\documentclass[12pt]{amsart}
\usepackage{a4wide,color}
\allowdisplaybreaks

\let\pa\partial

\let\eps\varepsilon

\newcommand{\R}{{\mathbb R}}
\newcommand{\T}{{\mathbb T}}

\let\ga=\gamma
\let\de=\delta
\let\eps=\varepsilon

\let\la=\lambda

\let\pa=\partial

\newtheorem{theorem}{Theorem}
\newtheorem{lemma}[theorem]{Lemma}
\newtheorem{proposition}[theorem]{Proposition}

\begin{document}

\title[Landau equation]{Exponential Time Decay Estimates for the Landau Equation on Torus}

\author[K.-C. Wu]{Kung-Chien Wu}
\address{Department of Pure Mathematics and Mathematical Statistics,
University of Cambridge, Wilberforce Road, Cambridge CB3 0WB, United Kingdom}
\email{kcw28@dpmms.cam.ac.uk; kungchienwu@gmail.com}

\date{\today}

\thanks{This paper is supervised by Cl\'{e}ment
Mouhot in Cambridge. This work is supported by the
Tsz-Tza Foundation in institute of mathematics, Academia Sinica,
Taipei, Taiwan.}

\begin{abstract}
We study the time decay estimates for the linearized Landau equation on torus when the initial
perturbation is not necessarily smooth.
Our result reveals the kinetic and fluid aspects of the equation. We design a
Picard-type iteration and Mixture lemma for constructing the increasingly regular kinetic like waves, they are carried by transport
equations and have exponential time decay rate. The fluid like waves are
constructed as part of the long-wave expansion in the spectrum of the Fourier
mode for the space variable and the time decay rate depends on the size of the domain. The Mixture lemma plays an important role in this paper, this lemma is parallel
to Boltzmann equation but the proof is more challenge.
\end{abstract}



\maketitle


\section{Introduction and Main Result}
\subsection{The Models}
The generalized Landau equation reads
\begin{equation}\label{in.1.a}
\left\{\begin{array}{l}
\displaystyle \pa_{t}F+\xi\cdot\nabla_{x}F=\frac{1}{\eps}Q(F,F)\,,
\\ \\
\displaystyle F(x,0,\xi)=F_{0}(x,\xi)\,,
\end{array}
\right.\end{equation}
with collision operator
$$
Q(F,G)=\nabla_{\xi}\cdot\Big[\int_{\R^{3}}\Phi(\xi-\xi_{*})
[F(\xi_{*})\nabla_{\xi}G(\xi)-G(\xi)\nabla_{\xi_{*}}F(\xi_{*})]d\xi_{*}\Big]\,.
$$
Here the positive constant $\eps$ is the Knudsen number, $F(t,x,\xi)\geq 0$ is the spatially periodic distribution function for the particles at time $t\geq0$, with spatial coordinates $x\in \T^{3}_{1}$, the 3-dimensional torus with unit size of each side and microscopic velocity $\xi\in\R^{3}$. The positive semi-definite matrix $\Phi(\xi)$ has the general form
$$
\Phi(\xi)=B(|\xi|)S(\xi)\,,
$$
where $B(|\xi|)=|\xi|^{\ga+2}$, $\ga\geq -3$, is a function depending on the nature of the interaction between the particles, and $S(\xi)$ is the 3 by 3 matrix
$$
S(\xi)=I_{3}-\frac{\xi\otimes \xi}{|\xi|^{2}}\,.
$$
The original Landau collision operator for the Coulombian interaction corresponds to the case $\ga=-3$.

In order to remove the parameter $\eps$ from the equation, we introduce the new scaled variables:
$$
\widetilde{x}=\frac{1}{\eps}x\,, \quad \widetilde{t}=\frac{1}{\eps}t\,,
$$
after dropping the tilde, the equation (\ref{in.1.a}) becomes
\begin{equation}\label{in.1.b}
\left\{\begin{array}{l}
\pa_{t}F+\xi\cdot\nabla_{x}F=Q(F,F)\,,\quad (x,t,\xi)\in \T^{3}_{1/\eps}\times\R^{+}\times\R^{3}
\\ \\
F(0,x,\xi)=F_{0}(x,\xi)\,,
\end{array}
\right.\end{equation}
where $\T^{3}_{1/\eps}$ means the 3-dimensional torus with size $1/\eps$ of each side.
The conservation of mass, momentum, as well as energy, can be formulated as
$$
\frac{d}{dt}\int_{\T^{3}_{1/\eps}}\int_{\R^{3}}\Big\{1,\xi,|\xi|^{2}\Big\}F(t,x,\xi)d\xi dx=0\,.
$$
As in the Boltzmann equation, it is well know that Maxwellians are steady states to the Landau equation (\ref{in.1.b}). We linearized the Landau equations (\ref{in.1.b}) around a global Maxwellian $M(\xi)$,
$$
M(\xi)=\frac{1}{(2\pi)^{3/2}}\exp\Big(\frac{-|\xi|^{2}}{2}\Big)\,,
$$
which the standard perturbation $f(x,t,\xi)$ to $M$ as
$$
F=M+M^{1/2}f\,.
$$
It is well-know that $Q(M,M)=0$, by expanding
$$
Q(M+M^{1/2}f, M+M^{1/2}f)=2Q(M,M^{1/2}f)+Q(M^{1/2}f,M^{1/2}f),
$$
dropping the nonlinear term, we can define the linearized collision Landau operator $L$ as
\begin{equation}
Lf=2M^{-1/2}Q(M,M^{1/2}f)\,.
\end{equation}
Similar to Boltzmann equation, we can decompose the linear collision operator $L$ as diffusion part and compact part.
$$
Lf= \widetilde{\Lambda}f +\widetilde{K}f\,,
$$
the diffusion part
$$
\widetilde{\Lambda}f =M^{-1/2}\nabla\cdot\big[(\vartheta M)\nabla(M^{-1/2}f)\big]\,,
$$
where the symmetric matrix $\vartheta(\xi)$ is
\begin{equation}\begin{array}{l}
\displaystyle \vartheta(\xi)=\int_{\R^{3}}\Phi(\xi-\xi_{*})M(\xi_{*})d\xi_{*}\,,
\end{array}\end{equation}
and the compact part
$$
\widetilde{K}f=\int_{\R^{3}}M^{-1/2}(\xi)M^{-1/2}(\xi_{*})\nabla_{\xi}\cdot\nabla_{\xi_{*}}Z(\xi,\xi_{*})f(\xi_{*})d\xi_{*}\,,
$$
where
$$
Z(\xi,\xi_{*})=M(\xi)M(\xi_{*})\Phi(\xi-\xi_{*})\,.
$$
The linearized Landau equation for $f(t,x,\xi)$ now takes the form
\begin{equation}\label{in.1.c}
\left\{\begin{array}{l}
\displaystyle\pa_{t}f+\xi\cdot\nabla_{x}f=Lf\,,
\\ \\
f(0,x,\xi)=I(x,\xi)\,,
\end{array}
\right.\end{equation}
here we define $f(x,t,\xi)=\mathbb{G}_{\eps}^{t}I(x,\xi)$, i.e. $\mathbb{G}_{\eps}^{t}$ is the solution operator (Green function) of the linearized Landau equation (\ref{in.1.c}). By assuming that initially $F_{0}(x,\xi)$ has the same total mass, momentum and total energy as the Maxwellian $M$, we can then rewrite the conservation laws as
\begin{equation}\label{in.1.d}
\int_{\T^{3}_{1/\eps}}\int_{\R^{3}}\Big\{1,\xi,|\xi|^{2}\Big\}M^{1/2}I(t,x,\xi)d\xi dx=0\,,
\end{equation}
this means the initial $I$ satisfies the zero mean conditions.

\subsection{Preliminaries}
Before the presentation of the properties of the collision operator $L$, let us define some notations in this paper.
For microscopic variable $\xi$, we shall use $L^{2}_{\xi}$
to denote the classical Hilbert space with norm
$$\Vert
f\Vert_{L^{2}_{\xi}}=\Big(\int_{\R^{3}} |f|^2d\xi\Big)^{1/2}\,,$$
the Sobolev space of functions with all its $s$-th partial derivatives in $L^{2}_{\xi}$
will be denoted by $H^{s}_{\xi}$. The $L^{2}_{\xi}$ inner product in $\R^{3}$ will be denoted by $\big<\cdot,\cdot\big>_{\xi}$.
For space variable $x$, we shall use $L^{2}_{x}$
to denote the classical Hilbert space with norm
$$\|f\|_{L^{2}_{x}}=\Big(\frac{1}{|\T^{3}_{1/\eps}|}\int_{\T^{3}_{1/\eps}}|f|^{2}dx\Big)^{1/2}\,,$$
the Sobolev space of functions with all its $s$-th partial derivatives in $L^{2}_{x}$
will be denoted by $H^{s}_{x}$. We denote the sup norm as
$$
\|f\|_{L^{\infty}_{x}}=\sup_{x\in \T^{3}_{1/\eps}}|f(x)|\,.
$$
For notational of simplicity, we denote $\big<\xi\big>^{s}=(1+|\xi|)^{s}$. In this paper, we need two projection operators,
\\
(i) For any $h\in L^{2}_{\xi}$, we define the projection to the space $\{M^{1/2}\}$ as
$$
p_{0}(h)=\big<h,M^{1/2}\big>_{\xi}M^{1/2}(\xi)\,.
$$
\\
(ii) For any vector function $u\in L^{2}_{\xi}$, we define the orthogonal projection to the vector $\xi$ as
$$
\mathbb{P}(\xi)u=\frac{u\cdot \xi}{|\xi|^{2}}\xi\,.
$$
First, we list some important propositions about the diffusion operator $\widetilde{\Lambda}$.
\begin{proposition} For any $\ga\geq -3$, we have
\\
\noindent{\rm(i)}
$$
\widetilde{\Lambda} f=\nabla_{\xi}\cdot\big[\vartheta\nabla_{\xi}f\big]+\nabla_{\xi}\cdot\big[\vartheta\xi\big]f-(\xi,\vartheta\xi )f\,.
$$
\noindent{\rm(ii)} The spectrum of $\vartheta(\xi)$ consists of a simple eigenvalue $\la_{1}(\xi)>0$ associated with the eigenvector $\xi$, and a double eigenvalue $\la_{2}(\xi)>0$ associated with eigenvector $\xi^{\perp}$. Moreover, there are constants $c_{1}$ and $c_{2}>0$ such that asymptotically, as $|\xi|\to \infty$, we have
$$
\la_{1}(\xi)=c_{1}\big<\xi\big>^{\ga},\quad \la_{2}(\xi)= c_{2}\big<\xi\big>^{\ga+2}\,,
$$
moreover
\begin{align*}
 (\xi,\vartheta\xi )&=\la_{1}(\xi)|\xi|^{2}\,, \\
(u,\vartheta u )&=\la_{1}(\xi)|\mathbb{P}(\xi)u|^{2}+\la_{2}(\xi)\big|\big[I_{3}-\mathbb{P}(\xi)\big]u\big|^{2}\,,
\end{align*}

\noindent{\rm(iii)}
For any $|\alpha|\geq 1$,
$$
|\pa_{\xi}^{\alpha}(\vartheta\xi)|+|\pa_{\xi}^{\alpha}\vartheta|\leq C_{\alpha}\big<\xi\big>^{\ga+2-|\alpha|}\,.
$$
and
$$
|\pa_{\xi}^{\alpha}\la_{1}(\xi)|\leq C_{\alpha}\big<\xi\big>^{\ga+1-|\alpha|}\,,\quad
|\pa_{\xi}^{\alpha}\la_{2}(\xi)|\leq C_{\alpha}\big<\xi\big>^{\ga+2-|\alpha|}\,.
$$
\noindent{\rm(iv) (Coercivity estimate)} There exists $c_{0}, \nu_{0}>0$ such that
\begin{align*}
\big<-\widetilde{\Lambda}(f), f\big>_{\xi}&\geq c_{0}\Big\{\|\big<\xi\big>^{\ga/2+1}f\|^{2}_{L^{2}_{\xi}}+\|\big<\xi\big>^{\ga/2}\mathbb{P}(\xi)\nabla_{\xi}f\|^{2}_{L^{2}_{\xi}} \\
&+\big\|\big<\xi\big>^{\ga/2+1}\big[I_{3}-\mathbb{P}(\xi)\big]\nabla_{\xi}f\big\|^{2}_{L^{2}_{\xi}}\Big\}
-\nu_{0}\big<M^{1/2}, f\big>_{\xi}^{2}
\,.
\end{align*}

\end{proposition}
It is easy to see that if we want to get coercivity estimate, we need modify the decomposition of $L$. Now, we define $L=\Lambda+K$, where
\begin{equation}\begin{array}{l}\label{cov}
\Lambda f=\widetilde{\Lambda}f-\nu_{0}p_{0}(f)\,,
\quad
Kf=\widetilde{K}f+\nu_{0}p_{0}(f)\,.
\end{array}\end{equation}
Under above definition, we can rewrite the coercivity estimate of Landau equation as
\begin{align*}
 \big<-\Lambda(f), f\big>_{\xi}&\geq c_{0}\Big\{\|\big<\xi\big>^{\ga/2+1}f\|^{2}_{L^{2}_{\xi}}+\|\big<\xi\big>^{\ga/2}\mathbb{P}(\xi)\nabla_{\xi}f\|^{2}_{L^{2}_{\xi}} \\
&+\big\|\big<\xi\big>^{\ga/2+1}\big[I_{3}-\mathbb{P}(\xi)\big]\nabla_{\xi}f\big\|^{2}_{L^{2}_{\xi}}\Big\}
\,.
\end{align*}

In order to estimate the Green function of the Landau equation in next section, we need to recall the spectrum $Spec(\eps k)$, $k\in\mathbb{Z}^{3}$, of the operator $-i\pi\eps\xi\cdot k +L$
\begin{proposition}\label{pr12}{\rm\cite{[Ellis]}}
For any $\ga\geq-2$, there exists $\de>0$ and $\tau=\tau(\de)>0$ such that
\\
\noindent{\rm(i)}\quad  For any $|\eps k|>\de$,
\begin{equation}\label{bgk.3.a}
\hbox{Spec}(\eps k)\subset\{z\in\mathbb{C} : Re(z)<-\tau\}\,.
\end{equation}
\\
\noindent{\rm(ii)}\quad  For any $|\eps k|<\de$, the spectrum within the region $\{z\in\mathbb{C} : Re(z)>-\tau\}$ consisting of exactly five eigenvalues $\{\sigma_{j}(\eps k )\}_{j=0}^{4}$,
\begin{equation}\label{bgk.3.b}
\hbox{Spec}(\eps k)\cap\{z\in\mathbb{C} : Re(z)>-\tau\}=
\{\sigma_{j}(\eps k )\}_{j=0}^{4}\,,
\end{equation}
and corresponding eigenvectors $\{e_{j}(\eps k )\}_{j=0}^{4}$,
where
\begin{equation}\begin{array}{l}\label{bgk.3.c}
\displaystyle \sigma_{j}(\eps k )=\sum_{n=1}^{3}a_{j,n}(i|\eps k|)^{n}+O(\eps k )^{4}\,,\quad a_{j,2}>0\,,
\\
\displaystyle e_{j}(\eps k )=\sum_{n=1}^{3}e_{j,n}(i|\eps k|)^{n}+O(\eps k )^{4}\,,\quad \big<e_{j},e_{k}\big>_{\xi}=\de_{jk}.
\end{array}
\end{equation}
\\
\noindent{\rm(iii)}\quad
\begin{equation}\label{bgk.3.d}
\displaystyle e^{(-i\pi\eps\xi\cdot k +L)t}f=\Pi_{\de}f+\chi_{\{|\eps k|<\de\}}\sum_{j=0}^{4}e^{\sigma_{j}(-\eps k)t}\big<e_{j}(\eps k ), f\big>_{\xi}e_{j}(\eps k )
\end{equation}
where $\|\Pi_{\de}\|_{L^{2}_{\xi}}=O(1)e^{-a(\tau)t}$, $a(\tau)>0$, $\chi_{\{\cdot\}}$ is the indicator function.
\end{proposition}
\subsection{Main Theorem and Method of Proof}
In this paper, we decompose the solution of the linearized Landau equation as kinetic part, fluid part and remainder part.
The kinetic aspect of the solution is described by the diffusive transport equation:
$$
\pa_{t}g+\xi\cdot\nabla_{x}g=\Lambda g\,.
$$
The operator $K$ is a smooth operator in $\xi$ variable, we use this smoothness property to design a
Picard-type iteration for constructing the increasingly regular kinetic like waves.

 The fluid behavior is studying by constructing the Green function with the Fourier series in space variable $x$:
$$
\mathbb{G}_{\eps}^{t}=\sum_{k\in\mathbb{Z}}\frac{1}{|\T^{3}_{1/\eps}|}e^{i\pi\eps k\cdot x+(-i\pi\eps\xi\cdot k +L)t}\,,
$$
here the Green function is viewed as an operator for function of $\xi$. This prompts the analysis of the spectrum of the operator $-i\pi\eps\xi\cdot k +L$. The spectrum includes five curves bifurcating from the origin. The origin is the multiple zero eigenvalues of $L$, the operator at $k=0$. The kernel of $L$ are the fluid variables. The fluid like waves are constructed from these curves near the origin.

With the kinetic like waves and fluid like waves constructed, the rest of the solution is of sufficient smoothness and it has exponential time decay rate.
\begin{theorem}\label{theorem1}
For any $\ga>-2$, $I\in L_{\xi}^{2}$ with compact support in $x$ and satisfies the zero mean conditions {\rm(\ref{in.1.d})}, then the solution
$$
\mathbb{G}_{\eps}^{t}I=\mathbb{G}_{\eps, K}^{t}I+\mathbb{G}_{\eps, F}^{t}I+\mathbb{G}_{\eps, R}^{t}I\,,
$$
consists of the kinetic part $\mathbb{G}_{\eps, K}^{t}I$: nonsmooth and time decay exponentially
\begin{equation}
\|\mathbb{G}_{\eps, K}^{t}I\|_{L^{2}_{x}L^{2}_{\xi}}=O(1)e^{-O(1)t}\,;
\end{equation}
the fluid part $\mathbb{G}_{\eps, F}^{t}I$: the time decay rate depends on the size of the domain, i.e. there exist $\de, \de_{0}, C_{0}>0$ such that
\\ \\
\noindent{\rm(i)} If $\eps>\de$,
\begin{equation}
\|\mathbb{G}_{\eps, F}^{t}I\|_{H^{s}_{x}L^{2}_{\xi}}=0\,,
\end{equation}
\\
\noindent{\rm(ii)} If $\de_{0}<\eps<\de$,
\begin{equation}
\|\mathbb{G}_{\eps, F}^{t}I\|_{H^{s}_{x}L^{2}_{\xi}}=O(1)e^{-O(1)\eps^{2}t}\,,
\end{equation}
\\
\noindent{\rm(iii)} If $0<\eps<\de_{0}$,
\begin{equation}
\|\mathbb{G}_{\eps, F}^{t}I\|_{H^{s}_{x}L^{2}_{\xi}}=O(1)\frac{C_{0}}{(1+t)^{3/2}}e^{-O(1)\eps^{2}t}\,;
\end{equation}
and the smooth remainder part $\mathbb{G}_{\eps, R}^{t}I$:
\begin{equation}
\|\mathbb{G}_{\eps, R}^{t}I\|_{H^{2}_{x}L^{2}_{\xi}}=O(1)e^{-O(1)t}\,.
\end{equation}
\end{theorem}

The Mixture Lemma plays an important role to construct the kinetic like waves, it states that the mixture of the two operators
$\mathbb{S}$ and $K$ in $\mathbb{M}^{t}_{j}$ (see section \ref{se4}) transports the regularity in the microscopic velocity
$\xi$ to the regularity of the space time $(x,t)$. This idea was introduced by Liu-Yu \cite{[LiuYu], [LiuYu2], [LiuYu3], [LiuYu1]}
to construct the Green function of the Boltzmann equation. In Liu-Yu's paper, the proof of Mixture lemma relies on the exact solution of
the damped transport equations. In this paper, we introduce a differential operator to avoid constructing explicit solution, this operator commutes with free transport operator and can transports the microscopic velocity regularity to space regularity, this idea can also apply to Boltzmann equation to simplify Liu-Yu's results. Actually, the Mixture Lemma is similar in
spirit as the well-known Averaging Lemma, see \cite{[Bouchut], [Golse], [Jabin]}. These
two lemmas have been introduced independently and used for different purposes.

The spectrum analysis of the Landau equation was introduced by Degond-Lemou \cite{[Degond]}, this result parallel to the Boltzmann case done by
 Ellis-Pinsky \cite{[Ellis]}, however, this paper is the first one give coercivity estimate of Landau equation (they called Poincare type inequality),
 and they also give some ideas to decompose the collection term as diffusion part and compact part, see (\ref{cov}). However, the Poincare type inequality
 is not enough in our analysis, we thank Mouhot \cite{[Mouhot]} and Guo \cite{[Guo]} give more shape coercivity estimates for the linearized
  Landau operator.

 The analysis based on spectrum has been carried out by many authors. The exponential time decay rates for the Boltzmann equation with hard potentials on
torus was firstly provided by Ukai \cite{[Ukai1]}. The time-asymptotic nonlinear stability was obtained in \cite{[Nishida3], [Ukai]}. Using the Nishida approach, \cite{[Nishida]} obtained the time-asymptotic
equivalent of Boltzmann solutions and Navier-Stokes solutions.
These works yield
the $L^{2}$ theory, since the Fourier transform is isometric in $L^2$.

In this paper, we need to restrict $\ga>-2$, there are two reasons about this restriction, first, we need spectrum gap analysis, but this only holds for
$\ga\geq -2$, if $\ga< -2$, the spectrum of the collision operator is not clear. The second reason is singular part of compact term $K_{s}$,
 the derivative of this operator need $L^{2}_{\xi}$ norm bound, and the coefficient need bounded by the cut off distance, this property holds only
 when $\ga>-2$, hence we need this restriction.

The pointwise description of the one-dimensional linearized Boltzmann equation with hard sphere was firstly provided by Liu-Yu \cite{[LiuYu]},
the fluid like waves can be constructed by both complex and spectrum analysis, it reveals the dissipative behavior
of the type of the Navier-Stokes equation as usually seems by the Chapman-Enskog expansion.
The kinetic like waves can be constructed by Picard-type iteration and Mixture lemma. In this paper,
we apply similar ideas to Landau equation on torus, we can also construct the kinetic and fluid like waves, which are both time
decay exponentially. Moreover, the decay rate of the fluid like waves depends on the size of the domain.

\subsection{Plain of the Paper}
The rest of the paper is organized as follows.  In section \ref{se4}, we design
Picard-type iteration and Mixture lemma for constructing the increasingly regular kinetic like waves. In section \ref{se2}, we construct the Green function of the linearized Landau equation on torus,
then use the long wave short wave decomposition and the spectrum analysis to obtain time decay rate, moreover,
 we improve the estimate of the fluid like waves.

\section{Kinetic Part}\label{se4}
In this section, we will use the kinetic decomposition to construct the kinetic part, and apply Mixture lemma to show the tail term has enough regularity in $x$.
\subsection{Picard iteration and Kinetic Decomposition}
We rewrite the linearized Landau equation (\ref{in.1.c}) as
\begin{equation}\label{bgk.4.a}
\left\{\begin{array}{l} \displaystyle\pa_{t}f+\xi\cdot\nabla_{x}f-\Lambda f=Kf\,,
\\ \\
\displaystyle f(x,0,\xi)=I(x,\xi)\,.
\\
\end{array}
\right.\end{equation}
The operator $K$ has some regularizing effects with respect to the microscopic variable $\xi$. We decompose it as $K=K_{s}+K_{r}$, $K_{s}\equiv K_{s,D}$ and $K_{r}\equiv K_{r,D}$:
\begin{equation}
\left\{\begin{array}{l}
\displaystyle K_{s}f=\int_{\R^{3}}\chi\Big(\frac{|\xi-\xi_{*}|}{D}\Big)M^{-1/2}(\xi)M^{-1/2}(\xi_{*})\nabla_{\xi}\cdot
\nabla_{\xi_{*}}Z(\xi,\xi_{*})f(\xi_{*})d\xi_{*}\,,
\\ \\
K_{r}f=Kf-K_{s}f\,,
\\ \\
\chi(r)=1 \quad \hbox{for} \quad r\in [-1,1]\,,
\\ \\
supp(\chi)\subset [-2,2]\,, \quad \chi\in C^{\infty}_{c}(\R)\,, \quad  \chi\geq 0\,.
\\
\end{array}
\right.\end{equation}
\begin{lemma} For $\ga>-2$, we have the following smoothing properties about the operator $K$ :
\begin{equation}\label{in.2.aa}
\|K_{s}f\|_{H^{1}_{\xi}}\leq D^{\ga+2}\|f\|_{L^{2}_{\xi}}\,, \quad
\|K_{r}f\|_{H^{s}_{\xi}}\leq C\|f\|_{L^{2}_{\xi}}\,.
\end{equation}
\end{lemma}
\noindent{\it Proof.} It is easy to see the regular part dominated by Guassian, so the estimate of regular part $K_{r}$ is obvious.
 For singular part $K_{s}$, we introduce the Hardy-Littlewood maximum function: for any $f\in L^{2}_{\xi}$
$$
\mathcal{M}(f)=\sup_{r>0}r^{-3}\int_{|\xi-\xi_{*}|<r}|f(\xi_{*})|d\xi_{*}\,,
$$
then $\pa_{\xi}K_{s}f$ can be bounded by Hardy-Littlewood maximum function $\mathcal{M}(f)$
\begin{align*}
|\pa_{\xi}K_{s}f(\xi)|&\leq C\int_{|\xi-\xi_{*}|<D}|\xi-\xi_{*}|^{(\ga+2)-3}|f(\xi_{*})|d\xi_{*} \\
&=C\sum_{n=0}^{\infty}\int_{2^{-(n+1)}D<|\xi-\xi_{*}|<2^{-n}D}|\xi-\xi_{*}|^{(\ga+2)-3}|f(\xi_{*})|d\xi_{*}\\
&\leq C\sum_{n=0}^{\infty}(2^{-n}D)^{\ga+2}(2^{-n}D)^{-3}\int_{|\xi-\xi_{*}|<2^{-n}D}|f(\xi_{*})|d\xi_{*}\\
&\leq C D^{\ga+2}\mathcal{M}(f)\,.
\end{align*}
By the Hardy-Littlewood-Wiener maximal theorem, we have
$$
\|\mathcal{M}(f)\|_{L^{2}_{\xi}}\leq C\|f\|_{L^{2}_{\xi}}\,,
$$
and hence
$$
\|\pa_{\xi}K_{s}f\|_{L^{2}_{\xi}}\leq CD^{\ga+2}\|f\|_{L^{2}_{\xi}}\,.
$$
\qed

We define the operator $g=\mathbb{S}^{t}g_{0}$ and $j=\mathbb{O}^{t}j_{0}$ as follows :
\begin{equation}
\left\{\begin{array}{l} \pa_{t}g +\xi\cdot\nabla_{x}g -\Lambda g =0\,,
\\ \\
g (x,0,\xi)=g_{0}(x,\xi)\,,
\\
\end{array}
\right.\end{equation}
and
\begin{equation}
\left\{\begin{array}{l} \pa_{t}j +\xi\cdot\nabla_{x}j -\Lambda j =K_{s}j\,,
\\ \\
j (x,0,\xi)=j_{0}(x,\xi)\,.
\\
\end{array}
\right.\end{equation}
We have the following standard estimates about diffusive transport equations:
\begin{lemma}\label{lemma1}
\begin{equation}
\|\mathbb{S}^{t}\|_{L^{2}_{x}L^{2}_{\xi}}\leq e^{-c_{0}t}\,,\quad
\|\mathbb{O}^{t}\|_{L^{2}_{x}L^{2}_{\xi}}\leq e^{-\frac{c_{0}}{2}t}\,.
\end{equation}
\end{lemma}
\noindent{\it Proof.} Note that $\ga>-2$, we have
\begin{equation}\begin{array}{l}
\displaystyle\frac{1}{2}\frac{d}{dt}\|g\|^{2}_{L^{2}_{x}L^{2}_{\xi}}=\frac{1}{|\T^{3}_{1/\eps}|}\int_{\R^{3}}\int_{ \T^{3}_{1/\eps}}\Lambda (g)g dxd\xi
\leq-c_{0}\|g\|^{2}_{L^{2}_{x}L^{2}_{\xi}}\,,
\end{array}\end{equation}
this proves $\|\mathbb{S}^{t}g_{0}\|_{L^{2}_{x}L^{2}_{\xi}}\leq e^{-c_{0}t}\|g_{0}\|^{2}_{L^{2}_{x}L^{2}_{\xi}}$. For $\mathbb{O}^{t}$, we have
\begin{equation}\begin{array}{l}
\displaystyle\frac{1}{2}\frac{d}{dt}\|j\|^{2}_{L^{2}_{x}L^{2}_{\xi}}=\frac{1}{|\T^{3}_{1/\eps}|}\int_{\R^{3}}\int_{ \T^{3}_{1/\eps}}\Lambda (j)j dxd\xi
\leq(-c_{0}+D^{\ga+2})\|j\|^{2}_{L^{2}_{x}L^{2}_{\xi}}\,,
\end{array}\end{equation}
we get our result if we choose $D$ sufficiently small.
\qed

Now, we design a Picard type iteration, which treat the regular part $K_{r}f$ as the source term. The $-1$ order approximation of the linearized Landau equation (\ref{bgk.4.a}) is the damped transport equation
\begin{equation}
\left\{\begin{array}{l} \pa_{t}h^{(-1)}+\xi\cdot\nabla_{x}h^{(-1)}-\Lambda h^{(-1)}=K_{s}h^{(-1)}\,,
\\ \\
h^{(-1)}(x,0,\xi)=I(x,\xi)\,.
\\
\end{array}
\right.\end{equation}
The difference $f-h^{(-1)}$ satisfies the equation
\begin{equation}
\left\{\begin{array}{l} \pa_{t}(f-h^{(-1)})+\xi\cdot\nabla_{x}(f-h^{(-1)})-\Lambda(f-h^{(-1)})=K(f-h^{(-1)})+K_{r}h^{(-1)}\,,
\\ \\
(f-h^{(-1)})(x,0,\xi)=0\,,
\\
\end{array}
\right.\end{equation}
this means the zero order approximation $h^{(0)}$ is
\begin{equation}
\left\{\begin{array}{l} \pa_{t}h^{(0)}+\xi\cdot\nabla_{x}h^{(0)}-\Lambda h^{(0)}=K_{r}h^{(-1)}\,,
\\ \\
h^{(0)}(x,0,\xi)=0\,,
\\
\end{array}
\right.\end{equation}
for this process, we can define the $k^{\hbox{th}}$ order approximation $h^{(k)}$, $k\geq 1$
\begin{equation}
\left\{\begin{array}{l} \pa_{t}h^{(k)}+\xi\cdot\nabla_{x}h^{(k)}-\Lambda h^{(k)}=Kh^{(k-1)}\,,
\\ \\
h^{(k)}(x,0,\xi)=0\,.
\\
\end{array}
\right.\end{equation}
This means the solution $f$ can be rewritten as
$$
 f=h^{(-1)}+h^{(0)}+h^{(1)}+\cdot\cdot\cdot\,.
$$
The following lemma gives the $L^{2}_{x}L^{2}_{\xi}$ estimate of $h^{(j)}$.
\begin{lemma}For $j\geq -1$, we have
\begin{equation}\begin{array}{l}\label{bgk.4.kq}
\|h^{(j)}\|_{L^{2}_{x}L^{2}_{\xi}}\leq t^{j+1}e^{-\frac{c_{0}}{2}t}\|I\|_{L^{2}_{x}L^{2}_{\xi}}\,.
\end{array}
\end{equation}
\end{lemma}
\noindent{\it Proof.} This lemma can be proven by induction. The case $j=-1$ immediately from Lemma \ref{lemma1}. Now, we prove the case $j=0$. By Duhamel principle,
\begin{equation}\begin{array}{l}
\displaystyle h^{(0)}(x,t,\xi)=\int_{0}^{t}\mathbb{S}^{t-s_{1}}K_{r}\mathbb{O}^{s_{1}}I(\cdot,s_{1})ds_{1}
\,.
\end{array}
\end{equation}
It is easy to see that
$$
\|h^{(0)}\|_{L^{2}_{x}L^{2}_{\xi}}\leq O(1) te^{-\frac{c_{0}t}{2}}\|I\|_{L^{2}_{x}L^{2}_{\xi}}\,.
$$
Assume it holds for $j$, then by the compactness property of $K$, we have
\begin{align*}
\|h^{(j+1)}\|_{L^{2}_{x}L^{2}_{\xi}}&=\Big\|\int_{0}^{t}\mathbb{S}^{t-s}(Kh^{(j)})(\cdot,s)ds\Big\|_{L^{2}_{x}L^{2}_{\xi}} \\
&\leq \int_{0}^{t}e^{-(t-s)}e^{-\frac{c_{0}s}{2}}s^{j+1}\|I\|_{L^{2}_{x}L^{2}_{\xi}}ds
\\
&\leq t^{j+2}e^{-\frac{c_{0}t}{2}}\|I\|_{L^{2}_{x}L^{2}_{\xi}}\,.
\end{align*}
\qed

Now, we can define the kinetic decomposition as
\begin{equation}\label{bgk.4.k}
\mathbb{G}_{\eps}^{t}I=\sum_{j=-1}^{4}h^{(j)}+\mathcal{R}=\mathbb{G}_{\eps,K}^{t}I+\mathcal{R}\,,
\end{equation}
then $\mathcal{R}$ satisfies the equation
\begin{equation}\label{bgk.4.l}
\left\{\begin{array}{l}
\pa_{t}\mathcal{R}+\xi\cdot\nabla_{x} \mathcal{R}=L\mathcal{R}+Kh^{(4)}\,,
\\ \\
\mathcal{R}(x,0,\xi)=0\,.
\\
\end{array}
\right.\end{equation}
By (\ref{bgk.4.l}), the regularity of $\mathcal{R}$ in $x$ is equivalent to the regularity of $h^{(4)}$ in $x$, in order to make sure the tail term $\mathcal{R}$ has good regularity, we need to check $h^{(4)}$ has regularity $H^{2}_{x}$. To proceed, we need the Mixture lemma, define the Mixture operator as follow:
\begin{equation}\label{bgk.4.r}
\mathbb{M}^{t}_{j}f_{0}=\int_{0}^{t}\int_{0}^{s_{1}}\cdot\cdot\cdot\int_{0}^{s_{2j-1}}\mathbb{S}^{t-s_{1}}K
\mathbb{S}^{s_{1}-s_{2}}K\mathbb{S}^{s_{2}-s_{3}}K\cdot\cdot\cdot\mathbb{S}^{s_{2j-1}-s_{2j}}K\mathbb{S}^{s_{2j}}
f_{0}ds_{2j}\cdot\cdot\cdot ds_{1}\,.
\end{equation}
Under this definition, we have
\begin{equation}\label{bgk.4.s}
h^{(4)}(x,t,\xi)=\int_{0}^{t}\mathbb{M}^{t-s_{0}}_{2}K_{r}\mathbb{O}^{s_{0}}I(\cdot,s_{0})ds_{0}\,.
\end{equation}
\begin{lemma}\label{mix}{\rm\cite{[Kuo], [LiuYu]}(Mixture Lemma)} For any $f_{0}\in L^{2}_{x}H^{2}_{\xi}$, $j=1,2$, we have
\begin{equation}\begin{array}{l}\label{bgk.4.t}
\|\pa^{j}_{x}\mathbb{M}^{t}_{j}f_{0}\|_{L^{2}_{x}L^{2}_{\xi}}\leq t^{j}e^{-\frac{2c_{0}t}{3}}\|f_{0}\|_{L^{2}_{x}H^{j}_{\xi}}\,.
\end{array}
\end{equation}
\end{lemma}
The proof of lemma \ref{mix} will be given in next subsection. We can estimate $h^{(4)}$ by using the Mixture lemma:
\begin{align}
 \|\pa^{2}_{x}h^{(4)}(\cdot,s)\|_{L^{2}_{x}L^{2}_{\xi}}&\leq\int_{0}^{s}(s-s_{0})^{2}e^{-\frac{2c_{0}(s-s_{0})}{3}}
\|\pa^{2}_{\xi}K_{r}\mathbb{O}^{s_{0}}I\|_{L^{2}_{x}L^{2}_{\xi}}ds_{0}\label{bgk.4.v} \\
&\leq e^{-\frac{c_{0}s}{2}}\int_{0}^{s}(s-s_{0})^{2}ds_{0}\|I\|_{L^{2}_{x}L^{2}_{\xi}}\nonumber\\
&\leq s^{3}e^{-\frac{c_{0}s}{2}}\|I\|_{L^{2}_{x}L^{2}_{\xi}}\nonumber\,.
\end{align}
Up to now, we have the kinetic part of Landau equation, and the tail term $\mathcal{R}$ has enough regularity in $x$, the construction of fluid part and remainder part will be given in section \ref{se2}.
\subsection{Proof of the Mixture Lemma}
The goal of this subsection is a short and direct proof of the Mixture lemma. The Mixture lemma was introduced by
 Liu-Yu \cite{[LiuYu], [LiuYu2], [LiuYu3], [LiuYu1]} to studying the Green function of the Boltzmann equation, and the proof relies on the
 exact solution of damped transport equation. However, it may difficult to construct the exact solution of $\mathbb{S}^{t}f_{0}$, in order to avoid this difficulty, we need to introduce a differential operator:
 $$
 \mathcal{D}_{t}=t\nabla_{x}+\nabla_{\xi}\,.
 $$
It is important that $\mathcal{D}_{t}$ commutes with free transport operator:
$$
[\mathcal{D}_{t},\pa_{t}+\xi\cdot\nabla_{x} ]=0\,.
$$
We have the following estimates about differential operator $\mathcal{D}_{t}$ and solution operator $\mathbb{S}^{t}$.
Now, we have the following estimates about $\mathcal{D}_{t}$ and $\mathbb{S}^{t}$.
\begin{lemma}\label{lemma-b}
For any $f_{0}\in L^{2}_{x}H^{j}_{\xi}$, $j=1,2$, there exists $\eta_{0}$ small enough such that
\begin{equation}\label{ml.1}
\left\{\begin{array}{l}
\displaystyle\|\mathcal{D}_{t}^{j}\mathbb{S}^{t}f_{0}\|_{L^{2}_{x}L^{2}_{\xi}}\leq e^{-(c_{0}-\eta_{0})t}\|f_{0}\|_{L^{2}_{x}H^{j}_{\xi}}\,,
\\ \\
\displaystyle\|\mathcal{D}_{t}\mathbb{S}^{t}f_{0}-\mathbb{S}^{t}\nabla_{\xi}f_{0}\|_{L^{2}_{x}L^{2}_{\xi}}\leq e^{-(c_{0}-\eta_{0})t}\|f_{0}\|_{L^{2}_{x}L^{2}_{\xi}}\,,
\\ \\
\displaystyle\|\mathcal{D}_{t}^{2}\mathbb{S}^{t}f_{0}-\mathcal{D}_{t}\mathbb{S}^{t}\nabla_{\xi}f_{0}\|_{L^{2}_{x}L^{2}_{\xi}}\leq e^{-(c_{0}-\eta_{0})t}\|f_{0}\|_{L^{2}_{x}H^{1}_{\xi}}\,.
\\
\end{array}
\right.\end{equation}
\end{lemma}
\textbf{Remark:} Although the integral operator $K$ has smoothing property, it only allows us differential with respect to $\xi$ once, if we calculate second order Mixture operator $\partial^{2}_{x}\mathbb{M}^{t}_{2}f_{0}$, the second derivative term $\pa^{2}_{\xi}K(\xi, \xi_{*})$ appears, so we need cancelation properties $(\ref{ml.1})_{2}$ and $(\ref{ml.1})_{3}$ to overcome this difficulty.

\noindent{\it Proof.}
Let $f^{(j)}=\mathcal{D}^{j}_{t}\mathbb{S}^{t}f_{0}$, $j=0,1,2$, then the energy estimate for $f^{(0)}$ gives
\begin{align*}
 \displaystyle\|f^{(0)}\|^{2}_{L^{2}_{x}L^{2}_{\xi}}&+c_{0}\int_{0}^{t}
\|\big<\xi\big>^{\ga/2+1}f^{(0)}\|^{2}_{L^{2}_{x}L^{2}_{\xi}}+\|\big<\xi\big>^{\ga/2}\mathbb{P}(\xi)\nabla_{\xi}f^{(0)}\|^{2}_{L^{2}_{x}L^{2}_{\xi}}\\
&+\big\|\big<\xi\big>^{\ga/2+1}\big[I_{3}-\mathbb{P}(\xi)\big]\nabla_{\xi}f^{(0)}\big\|^{2}_{L^{2}_{x}L^{2}_{\xi}}ds
\leq\|f_{0}\|^{2}_{L^{2}_{x}L^{2}_{\xi}}
\,.
\end{align*}
We study the operator $\mathcal{D}_{t}$ componentwise, on the other hand, define $\mathcal{D}_{it}=t\pa_{x^{i}}+\pa_{\xi^{i}}$, $i=1,2,3$. We can check the commutator
\begin{align*}
 \displaystyle[\mathcal{D}_{it},\Lambda ]h&=\nabla_{\xi}\cdot\big[(\pa_{\xi^{i}}\vartheta)\nabla_{\xi}h\big]+\nabla_{\xi}\cdot
\big[\pa_{\xi^{i}}(\vartheta\xi)\big]h-
\big[\pa_{\xi^{i}}(\xi,\vartheta\xi )\big]h\\
&-\nu_{0}\big<M^{1/2},h\big>_{\xi}\big(\pa_{\xi^{i}}M^{1/2}\big)+\nu_{0}\big<M^{1/2},\pa_{\xi^{i}}h\big>_{\xi}M^{1/2}\\
&=\sum_{l=1}^{5}B^{l}_{i}(h)\,.
\end{align*}
Denote
$$
\displaystyle[\mathcal{D}_{t},\Lambda ]h=\sum_{l=1}^{5}B^{l}(h)\,,\quad B^{l}=(B^{l}_{1},B^{l}_{2},B^{l}_{3}) \,,
$$
then $f^{(1)}$ solves the equation
$$
\pa_{t}f^{(1)}+\xi\cdot\nabla_{x}f^{(1)}=\Lambda f^{(1)}+[\mathcal{D}_{t},\Lambda ]f^{(0)}\,,
$$
by energy estimate of $f^{(1)}$, we have
\begin{align*}
\frac{1}{2}\frac{d}{dt}\|f^{(1)}\|^{2}_{L^{2}_{x}L^{2}_{\xi}}
=\frac{1}{|\T^{3}_{1/\eps}|}\int_{\R^{3}}\int_{ \T^{3}_{1/\eps}}\Lambda (f^{(1)})f^{(1)} dxd\xi
+\frac{1}{|\T^{3}_{1/\eps}|}\int_{\R^{3}}\int_{ \T^{3}_{1/\eps}}\sum_{l=1}^{5}B^{l}(f^{(0)})f^{(1)} dxd\xi\,.
\end{align*}
Now, we estimate $B^{1},\cdot\cdot\cdot, B^{5}$ separately. For $B^{2}$, note that $|\nabla_{\xi}\cdot\big[\pa_{\xi^{i}}(\vartheta\xi)\big]|\leq\big<\xi\big>^{\ga}$, hence
\begin{align*}
 \Big|\frac{1}{|\T^{3}_{1/\eps}|}\int_{\R^{3}}\int_{ \T^{3}_{1/\eps}}B^{2}(f^{(0)}) \cdot f^{(1)} dxd\xi\Big|
\leq C\|\big<\xi\big>^{\ga/2}f^{(0)}\|^{2}_{L^{2}_{x}L^{2}_{\xi}}+\de_{1}
 \|\big<\xi\big>^{\ga/2}f^{(1)}\|^{2}_{L^{2}_{x}L^{2}_{\xi}}\,.
\end{align*}
For $B^{3}$, we have $|\pa_{\xi^{i}}(\xi,\vartheta\xi )|\leq\big<\xi\big>^{\ga+2}$, this means
\begin{align*}
\Big|\frac{1}{|\T^{3}_{1/\eps}|}\int_{\R^{3}}\int_{ \T^{3}_{1/\eps}}B^{3}(f^{(0)}) \cdot f^{(1)} dxd\xi\Big|
\leq C\|\big<\xi\big>^{\ga/2+1}f^{(0)}\|^{2}_{L^{2}_{x}L^{2}_{\xi}}+\de_{2} \|\big<\xi\big>^{\ga/2+1}f^{(1)}\|^{2}_{L^{2}_{x}L^{2}_{\xi}}\,,
\end{align*}
and for $B^{4}+B^{5}$,
\begin{align*}
\Big|\frac{1}{|\T^{3}_{1/\eps}|}\int_{\R^{3}}\int_{ \T^{3}_{1/\eps}}\big(B^{4}+B^{5}\big)(f^{(0)}) \cdot f^{(1)} dxd\xi\Big|
\leq C\|f^{(0)}\|^{2}_{L^{2}_{x}L^{2}_{\xi}}+\de_{3} \|f^{(1)}\|^{2}_{L^{2}_{x}L^{2}_{\xi}}\,.
\end{align*}
Finally, for $B^{1}$, we need to decompose the microscopic velocity $\xi$ near origin and away from origin. Notice that for a
smooth cut-off function $\chi(\xi)=1$ for $|\xi|\leq1$, $\chi(\xi)=0$ for $|\xi|\geq2$, going back to componentwise expression $B^{1}_{i}$, we have
\begin{align*}
\int_{\R^{3}}B^{1}_{i}(f^{(0)})  f^{(1)} d\xi=\int_{|\xi|\leq 2}\chi B^{1}_{i}(f^{(0)})  f^{(1)} d\xi
+\int_{|\xi|\geq 1}(1-\chi)B^{1}_{i}(f^{(0)})  f^{(1)} d\xi\,.
\end{align*}
For $|\xi|$ small, we have
\begin{align*}
  &\phantom{xx}{}\int_{|\xi|\leq 2}\chi B^{1}_{i}(f^{(0)})  f^{(1)} d\xi=\int_{|\xi|\leq 2}-\nabla_{\xi}(\chi f^{(1)})(\pa_{\xi^{i}}\vartheta)\nabla_{\xi}f^{(0)}d\xi\\
&\leq C \Big(\int_{|\xi|\leq 2}|\nabla_{\xi}f^{(0)}|^{2}d\xi \Big)+
\de_{4}\Big(\int_{|\xi|\leq 2}|\nabla_{\xi}f^{(1)}|^{2}+|f^{(1)}|^{2}d\xi \Big)\,.
\end{align*}
For $|\xi|$ large, we have
\begin{align*}
  &\phantom{xx}{}\int_{|\xi|\ge 1}(1-\chi) B^{1}_{i}(f^{(0)})  f^{(1)} d\xi=\int_{|\xi|\ge 1}-\nabla_{\xi}((1-\chi) f^{(1)})(\pa_{\xi^{i}}\vartheta)\nabla_{\xi}f^{(0)}d\xi\\
&\leq  C\|\big<\xi\big>^{\ga/2}\nabla_{\xi}f^{(0)}\|_{L^{2}_{\xi}}\|\big<\xi\big>^{\ga/2+1}f^{(1)}\|_{L^{2}_{\xi}}
+\int_{|\xi|\ge 1}\pa_{\xi^{i}}(1-\chi)\nabla_{\xi} f^{(1)}\vartheta\nabla_{\xi}f^{(0)}d\xi\\
&+ \int_{|\xi|\ge 1}(1-\chi)\nabla_{\xi} \pa_{\xi^{i}}f^{(1)}\vartheta\nabla_{\xi}f^{(0)}d\xi
 +\int_{|\xi|\ge 1}(1-\chi)\nabla_{\xi} f^{(1)}\vartheta\nabla_{\xi}\pa_{\xi^{i}}f^{(0)}d\xi\,.
\end{align*}
We only need to estimate the last two integral,
\begin{align*}
  &\phantom{xx}{} \int_{|\xi|\ge 1}(1-\chi)\nabla_{\xi} \pa_{\xi^{i}}f^{(1)}\vartheta\nabla_{\xi}f^{(0)}d\xi
 +\int_{|\xi|\ge 1}(1-\chi)\nabla_{\xi} f^{(1)}\vartheta\nabla_{\xi}\pa_{\xi^{i}}f^{(0)}d\xi\\
&=\int_{|\xi|\ge 1}(1-\chi)\Big[ \la_{1}\Big(\mathbb{P}(\xi)\nabla_{\xi}\pa_{\xi^{i}}f^{(1)},\mathbb{P}(\xi)\nabla_{\xi}f^{(0)}\Big)
+\la_{2}\Big([I_{3}-\mathbb{P}(\xi)]\nabla_{\xi}\pa_{\xi^{i}}f^{(1)},[I_{3}-\mathbb{P}(\xi)]\nabla_{\xi}f^{(0)}\Big)\Big]d\xi
\\
&+\int_{|\xi|\ge 1}(1-\chi)\Big[ \la_{1}\Big(\mathbb{P}(\xi)\nabla_{\xi}f^{(1)},\mathbb{P}(\xi)\nabla_{\xi}\pa_{\xi^{i}}f^{(0)}\Big)
+ \la_{2}\Big([I_{3}-\mathbb{P}(\xi)]\nabla_{\xi}f^{(1)},[I_{3}-\mathbb{P}(\xi)]\nabla_{\xi}\pa_{\xi^{i}}f^{(0)}\Big)\Big]d\xi
\\
&=-\int_{|\xi|\ge 1}\pa_{\xi^{i}}(1-\chi)\nabla_{\xi} f^{(1)}\vartheta\nabla_{\xi}f^{(0)}d\xi
-\int_{|\xi|\ge 1}(1-\chi)\pa_{\xi^{i}}\la_{1}(\xi)\Big(\mathbb{P}(\xi)\nabla_{\xi}f^{(1)},\mathbb{P}(\xi)\nabla_{\xi}f^{(0)}\Big)d\xi
\\
&-\int_{|\xi|\ge 1}(1-\chi)\pa_{\xi^{i}}\la_{2}(\xi)\Big([I_{3}-\mathbb{P}(\xi)]\nabla_{\xi}f^{(1)},[I_{3}-\mathbb{P}(\xi)]\nabla_{\xi}f^{(0)}\Big)d\xi
\\
&-\int_{|\xi|\ge 1}(1-\chi)\big[\la_{1}(\xi)-\la_{2}(\xi)\big]\big([I_{3}-\mathbb{P}(\xi)]\nabla_{\xi}f^{(1)},e_{i}\big)
\frac{\big(\nabla_{\xi}f^{(0)},\xi\big)}{|\xi|^{2}}d\xi
\\
&-\int_{|\xi|\ge 1}(1-\chi)\big[\la_{1}(\xi)-\la_{2}(\xi)\big]\big([I_{3}-\mathbb{P}(\xi)]\nabla_{\xi}f^{(0)},e_{i}\big)
\frac{\big(\nabla_{\xi}f^{(1)},\xi\big)}{|\xi|^{2}}d\xi\,.
\end{align*}
Hence we have
\begin{align*}
 &\phantom{xx}{}\int_{|\xi|\ge 1}(1-\chi) B^{1}_{i}(f^{(0)})  f^{(1)} d\xi\\
&\leq C\Big(\|\big<\xi\big>^{\ga/2}\nabla_{\xi}f^{(0)}\|^{2}_{L^{2}_{\xi}}+
\|\big<\xi\big>^{\ga/2}\mathbb{P}(\xi)\nabla_{\xi}f^{(0)}\|^{2}_{L^{2}_{\xi}}
+\|\big<\xi\big>^{\ga/2+1}[I_{3}-\mathbb{P}(\xi)]\nabla_{\xi}f^{(0)}\|^{2}_{L^{2}_{\xi}}\Big)\\
&+ \de_{5}\Big(\|\big<\xi\big>^{\ga/2+1}f^{(1)}\|^{2}_{L^{2}_{\xi}}+\|\big<\xi\big>^{\ga/2}\mathbb{P}(\xi)\nabla_{\xi}f^{(1)}\|^{2}_{L^{2}_{\xi}}
+\|\big<\xi\big>^{\ga/2+1}[I_{3}-\mathbb{P}(\xi)]\nabla_{\xi}f^{(1)}\|^{2}_{L^{2}_{\xi}}\Big)\,.
\end{align*}
Summing up above estimates for $B_{1}^{1}, B_{2}^{1}$ and $B_{3}^{1}$, we have
\begin{align*}
 &\phantom{xx}{}\Big|\frac{1}{|\T^{3}_{1/\eps}|}\int_{\R^{3}}\int_{ \T^{3}_{1/\eps}}B^{1}(f^{(0)}) \cdot f^{(1)} dxd\xi\Big|\\
&\leq C\Big(\|\big<\xi\big>^{\ga/2}\mathbb{P}(\xi)\nabla_{\xi}f^{(0)}\|^{2}_{L^{2}_{x}L^{2}_{\xi}}+\|\big<\xi\big>^{\ga/2+1}\big[I_{3}-\mathbb{P}(\xi)\big]
\nabla_{\xi}f^{(0)}\|^{2}_{L^{2}_{x}L^{2}_{\xi}}\Big)\\
&+\widetilde{\de}\Big(\|\big<\xi\big>^{\ga/2+1}f^{(1)}\|^{2}_{L^{2}_{\xi}}+
\|\big<\xi\big>^{\ga/2}\mathbb{P}(\xi)\nabla_{\xi}f^{(1)}\|^{2}_{L^{2}_{x}L^{2}_{\xi}}+\|\big<\xi\big>^{\ga/2+1}\big[I_{3}-\mathbb{P}(\xi)\big]
\nabla_{\xi}f^{(1)}\|^{2}_{L^{2}_{x}L^{2}_{\xi}}\Big)\,,
\end{align*}
where $\widetilde{\de}$ depend on $\de_{4},\de_{5}$. Choose $\de_{1}, \de_{2}$, $\de_{3}$ and $\widetilde{\de}$ small enough, then there exists $\eta_{0}$ small enough such taht
\begin{equation}
\|f^{(1)}\|^{2}_{L^{2}_{x}L^{2}_{\xi}}\leq -(c_{0}-\eta_{0})\int_{0}^{t}\|f^{(1)}\|^{2}_{L^{2}_{x}L^{2}_{\xi}}ds
+\|f_{0}\|^{2}_{L^{2}_{x}L^{2}_{\xi}}+\|\nabla_{\xi}f_{0}\|^{2}_{L^{2}_{x}L^{2}_{\xi}}\,,
\end{equation}
and hence we obtain
$$
\|\mathcal{D}_{t}\mathbb{S}^{t}f_{0}\|_{L^{2}_{x}L^{2}_{\xi}}\leq e^{-(c_{0}-\eta_{0})t}\|f_{0}\|_{L^{2}_{x}H^{1}_{\xi}}\,.
$$
For second order estimate, we have
\begin{equation}
\pa_{t}f^{(2)}+\xi\cdot\nabla_{x}f^{(2)}=\Lambda f^{(2)}+2[\mathcal{D}_{t},\Lambda ]f^{(1)}+\big[\mathcal{D}_{t},[\mathcal{D}_{t},\Lambda] \big]f^{(0)}\,,
\end{equation}
the second order commutator $\big[\mathcal{D}_{t},[\mathcal{D}_{t},\Lambda] \big]h=\{A_{ij}(h)\}_{i,j=1}^{3}$ is a matrix, where
\begin{align*}
A_{ij}(h)&=\nabla_{\xi}\cdot\big[(\pa^{2}_{\xi^{i}\xi^{j}}\vartheta)\nabla_{\xi}h\big]+\nabla_{\xi}\cdot
\big[\pa^{2}_{\xi^{i}\xi^{j}}(\vartheta\xi)\big]h-
\big[\pa^{2}_{\xi^{i}\xi^{j}}(\xi,\vartheta\xi )\big]h\\
&-\nu_{0}\big<M^{1/2},h\big>_{\xi}\big(\pa^{2}_{\xi^{i}\xi^{j}}M^{1/2}\big)+\nu_{0}\big<M^{1/2},\pa_{\xi^{i}}h\big>_{\xi}(\pa_{\xi^{j}}M^{1/2})\\
&-\nu_{0}\big<M^{1/2},\pa^{2}_{\xi^{i}\xi^{j}}h\big>_{\xi}M^{1/2}+\nu_{0}\big<M^{1/2},\pa_{\xi^{j}}h\big>_{\xi}(\pa_{\xi^{i}}M^{1/2})\,,
\end{align*}
similar argument can get our result and hence we omit the detail, this proves the inequality $(\ref{ml.1})_{1}$.
For $(\ref{ml.1})_{2}$ and $(\ref{ml.1})_{3}$, we can define $g_{0}=\nabla_{\xi}f_{0}$ and $g^{(j)}=\mathcal{D}_{t}\mathbb{S}^{t}g_{0}$, then $g^{(0)}$
 and $g^{(1)}$ satisfy the following equation respectively
\begin{equation}
\left\{\begin{array}{l}
\pa_{t}g^{(0)}+\xi\cdot\nabla_{x}g^{(0)}=\Lambda g^{(0)}\,,
\\ \\
\displaystyle g^{(0)}(x,0,\xi)=\nabla_{\xi}f_{0}\,,
\\
\end{array}
\right.\end{equation}
and
\begin{equation}
\left\{\begin{array}{l}
\pa_{t}g^{(1)}+\xi\cdot\nabla_{x}g^{(1)}=\Lambda g^{(1)}+[\mathcal{D}_{t},\Lambda ]g^{(0)}\,,
\\ \\
\displaystyle g^{(1)}(x,0,\xi)=\nabla^{2}_{\xi}f_{0}\,.
\\
\end{array}
\right.\end{equation}
We can define $u^{(j)}=f^{(j)}-g^{(j-1)}$, $j=1,2$. Then $u^{(1)}=\mathcal{D}_{t}\mathbb{S}^{t}f_{0}-\mathbb{S}^{t}\nabla_{\xi}f_{0}$
and $u^{(2)}=\mathcal{D}_{t}^{2}\mathbb{S}^{t}f_{0}-\mathcal{D}_{t}\mathbb{S}^{t}\nabla_{\xi}f_{0}$. For $(\ref{ml.1})_{2}$, note that
$u^{(1)}$ solves the equation
\begin{equation}
\left\{\begin{array}{l}
\pa_{t}u^{(1)}+\xi\cdot\nabla_{x}u^{(1)}=\Lambda u^{(1)}+[\mathcal{D}_{t},\Lambda ]f^{(0)}\,,
\\ \\
\displaystyle u^{(1)}(x,0,\xi)=0\,,
\\
\end{array}
\right.\end{equation}
the energy estimate gives
$$
\|u^{(1)}\|^{2}_{L^{2}_{x}L^{2}_{\xi}}\leq -(c_{0}-\eta_{0})\int_{0}^{t}\|u^{(1)}\|^{2}_{L^{2}_{x}L^{2}_{\xi}}ds+\|f_{0}\|^{2}_{L^{2}_{x}L^{2}_{\xi}}\,,
$$
this complete the proof of $(\ref{ml.1})_{2}$. For $(\ref{ml.1})_{3}$, $u^{(2)}$ solves the equation
\begin{equation}
\left\{\begin{array}{l}
\pa_{t}u^{(2)}+\xi\cdot\nabla_{x}u^{(2)}-\Lambda u^{(2)}
=2[\mathcal{D}_{t},\Lambda ]f^{(1)}+\big[\mathcal{D}_{t},[\mathcal{D}_{t},\Lambda] \big]f^{(0)}-[\mathcal{D}_{t},\Lambda ]g^{(0)}\,,
\\ \\
\displaystyle u^{(2)}(x,0,\xi)=0\,,
\\
\end{array}
\right.\end{equation}
similar argument can get our result.
\qed
\bigskip

\noindent{\it Proof of the Mixture Lemma.} For $j=1$, we can write down $\pa^{1}_{x}\mathbb{M}^{t}_{1}f_{0}$ as follows:
\begin{align*}
  \pa^{1}_{x}\mathbb{M}^{t}_{1}f_{0}(x,\xi)&=
  \int_{0}^{t}\int_{0}^{s_{1}}
\pa_{x}\mathbb{S}^{t-s_{1}}K
\mathbb{S}^{s_{1}-s_{2}}K\mathbb{S}^{s_{2}}f_{0}ds_{2}ds_{1}\\
&=\int_{0}^{t}\int_{0}^{s_{1}}\int_{\R^{6}}\pa_{x}\mathbb{S}^{t-s_{1}}W(\xi,\xi_{1})\mathbb{S}^{s_{1}-s_{2}}W(\xi_{1},\xi_{2})
\mathbb{S}^{s_{2}}f_{0}d\xi_{2}d\xi_{1}ds_{2}ds_{1}\,.
\end{align*}
Note that $[\pa_{x}, \mathbb{S}^{t}]=0$, $[\pa_{x},W]=0$, we can change the order of $(\pa_{x}, \mathbb{S}^{t})$ and $(\pa_{x},W)$. In order to get time integrability, we can rewrite $\pa^{1}_{x}\mathbb{M}^{t}_{1}f_{0}(x,\xi)$ as
\begin{align*}
\pa^{1}_{x}\mathbb{M}^{t}_{1}f_{0}(x,\xi)&=\int_{0}^{t}\int_{0}^{s_{1}}\int_{\R^{6}}\frac{s_{1}-s_{2}}{s_{1}}\mathbb{S}^{t-s_{1}}W(\xi,\xi_{1})\pa_{x}\mathbb{S}^{s_{1}-s_{2}}W(\xi_{1},\xi_{2})
\mathbb{S}^{s_{2}}f_{0}d\xi_{2}d\xi_{1}ds_{2}ds_{1}\\
  &+\int_{0}^{t}\int_{0}^{s_{1}}\int_{\R^{6}}\frac{s_{2}}{s_{1}}\mathbb{S}^{t-s_{1}}W(\xi,\xi_{1})\mathbb{S}^{s_{1}-s_{2}}W(\xi_{1},\xi_{2})
\pa_{x}\mathbb{S}^{s_{2}}f_{0}d\xi_{2}d\xi_{1}ds_{2}ds_{1}\,.
\end{align*}
Using the fact $t\nabla_{x}=\mathcal{D}_{t}-\nabla_{\xi}$, we have
\begin{align*}
 \pa^{1}_{x}\mathbb{M}^{t}_{1}f_{0}(x,\xi)&=\int_{0}^{t}\int_{0}^{s_{1}}\int_{\R^{6}}\frac{1}{s_{1}}\mathbb{S}^{t-s_{1}}W(\xi,\xi_{1})
\big(\mathcal{D}_{s_{1}-s_{2}}-\nabla_{\xi_{1}}\big)\mathbb{S}^{s_{1}-s_{2}}W(\xi_{1},\xi_{2})
\mathbb{S}^{s_{2}}f_{0}d\xi_{2}d\xi_{1}ds_{2}ds_{1} \\
&+\int_{0}^{t}\int_{0}^{s_{1}}\int_{\R^{6}}\frac{1}{s_{1}}\mathbb{S}^{t-s_{1}}W(\xi,\xi_{1})\mathbb{S}^{s_{1}-s_{2}}W(\xi_{1},\xi_{2})
\big(\mathcal{D}_{s_{2}}-\nabla_{\xi_{2}}\big)\mathbb{S}^{s_{2}}f_{0}d\xi_{2}d\xi_{1}ds_{2}ds_{1}\,.
\end{align*}
By $(\ref{ml.1})_{1}$ and integration by parts, we have
\begin{align*}
 \|\partial^{1}_{x}\mathbb{M}^{t}_{1}f_{0}\|_{L^{2}_{x}L^{2}_{\xi}}&\leq e^{-\frac{2c_{0}t}{3}}\big(\|f_{0}\|_{L^{2}_{x}L^{2}_{\xi}}
+\|\partial_{\xi}^{1}f_{0}\|_{L^{2}_{x}L^{2}_{\xi}}\big)\Big(\int_{0}^{t}\int_{0}^{s_{1}}\frac{1}{s_{1}}ds_{2}ds_{1}\Big) \\
&=te^{-\frac{2c_{0}t}{3}}\big(\|f_{0}\|_{L^{2}_{x}L^{2}_{\xi}}
+\|\partial_{\xi}^{1}f_{0}\|_{L^{2}_{x}L^{2}_{\xi}}\big)\,.
\end{align*}
This proves the case $j=1$. For $j=2$, we can write down $\partial^{2}_{x}\mathbb{M}^{t}_{2}f_{0}$ as:
\begin{align*}
  &\phantom{xx}{} \partial^{2}_{x}\mathbb{M}^{t}_{2}f_{0}(x,\xi)\\
  &=\int_{\T}\int_{\R^{6\times 2}}\frac{s_{1}s_{3}}{s_{1}s_{3}}\partial^{2}_{x}
\Big[\mathbb{S}^{t-s_{1}}W(\xi,\xi_{1})
\mathbb{S}^{s_{1}-s_{2}}W(\xi_{1},\xi_{2})
\mathbb{S}^{s_{2}-s_{3}}W(\xi_{2},\xi_{3})\mathbb{S}^{s_{3}-s_{4}}W(\xi_{3},\xi_{4})
\mathbb{S}^{s_{4}}f_{0}\Big]d\Xi dS\,,
\end{align*}
where
$$
dS=ds_{1}\cdot\cdot\cdot ds_{4},\quad d\Xi=d\xi_{1}\cdot\cdot\cdot d\xi_{4}\,,\quad \T=[0,t]\times[0,s_{1}]\times\cdot\cdot\cdot\times[0,s_{3}]\,.
$$
In order to get time integrability, we need decompose $s_{1}s_{3}$ as
$$s_{1}s_{3}=[(s_{1}-s_{2})+(s_{2}-s_{3})+(s_{3}-s_{4})+s_{4}][(s_{3}-s_{4})+s_{4}]\,.$$
We then have
\begin{equation*}\begin{array}{l}
\displaystyle\partial^{2}_{x}\mathbb{M}^{t}_{2}f_{0}(x,\xi)=\int_{\T}\int_{\R^{6\times 2}}\frac{1}{s_{1}s_{3}}(J_{1}f_{0}+J_{2}f_{0})d\Xi dS\,,
\end{array}\end{equation*}
where $J_{1}$ collects all the terms that each $W$ differential with respect to $\xi$ at most once, and $J_{2}$ collects all the terms that one of $W$ differential with respect to $\xi$ twice. More precisely,
\begin{align}
 J_{2}f_{0}
&=\mathbb{S}^{t-s_{1}}W(\xi,\xi_{1})
\mathbb{S}^{s_{1}-s_{2}}W(\xi_{1},\xi_{2})
\big[\mathcal{D}_{s_{2}-s_{3}}\mathbb{S}^{s_{2}-s_{3}}\big]\big[\nabla_{\xi_{3}}W(\xi_{2},\xi_{3})\big]\mathbb{S}^{s_{3}-s_{4}}W(\xi_{3},\xi_{4})
\mathbb{S}^{s_{4}}f_{0}\nonumber
\\
& +\mathbb{S}^{t-s_{1}}W(\xi,\xi_{1})
\mathbb{S}^{s_{1}-s_{2}}W(\xi_{1},\xi_{2})
\mathbb{S}^{s_{2}-s_{3}}\big[\nabla^{2}_{\xi_{3}}W(\xi_{2},\xi_{3})\big]\mathbb{S}^{s_{3}-s_{4}}W(\xi_{3},\xi_{4})
\mathbb{S}^{s_{4}}f_{0}\nonumber
\\
&+\mathbb{S}^{t-s_{1}}W(\xi,\xi_{1})
\mathbb{S}^{s_{1}-s_{2}}W(\xi_{1},\xi_{2})
\mathbb{S}^{s_{2}-s_{3}}W(\xi_{2},\xi_{3})\big[\mathcal{D}_{s_{3}-s_{4}}\mathbb{S}^{s_{3}-s_{4}}\big]\big[\nabla_{\xi_{4}}W(\xi_{3},\xi_{4})\big]
\mathbb{S}^{s_{4}}f_{0}\label{bot.6.f}
\\
&+\mathbb{S}^{t-s_{1}}W(\xi,\xi_{1})
\mathbb{S}^{s_{1}-s_{2}}W(\xi_{1},\xi_{2})
\mathbb{S}^{s_{2}-s_{3}}W(\xi_{2},\xi_{3})\mathbb{S}^{s_{3}-s_{4}}\big[\nabla^{2}_{\xi_{4}}W(\xi_{3},\xi_{4})\big]
\mathbb{S}^{s_{4}}f_{0}\nonumber
\\
&+\mathbb{S}^{t-s_{1}}W(\xi,\xi_{1})
\mathbb{S}^{s_{1}-s_{2}}W(\xi_{1},\xi_{2})
\mathbb{S}^{s_{2}-s_{3}}W(\xi_{2},\xi_{3})\big[\mathcal{D}^{2}_{s_{3}-s_{4}}\mathbb{S}^{s_{3}-s_{4}}\big]W(\xi_{3},\xi_{4})
\mathbb{S}^{s_{4}}f_{0}\nonumber
\\
&+\mathbb{S}^{t-s_{1}}W(\xi,\xi_{1})
\mathbb{S}^{s_{1}-s_{2}}W(\xi_{1},\xi_{2})
\mathbb{S}^{s_{2}-s_{3}}W(\xi_{2},\xi_{3})\big[\mathcal{D}_{s_{3}-s_{4}}\mathbb{S}^{s_{3}-s_{4}}\big]\big[\nabla_{\xi_{4}}W(\xi_{3},\xi_{4})\big]
\mathbb{S}^{s_{4}}f_{0}\nonumber\,.
\end{align}
The estimate of $J_{1}$ is similar to $j=1$,
$$
\Big\|\int_{\R^{6\times 2}}J_{1}f_{0}d\Xi\Big\|_{L^{2}_{x}L^{2}_{\xi}}\leq e^{-\frac{2c_{0}t}{3}}\|f_{0}\|_{L^{2}_{x}H^{2}_{\xi}}\,.
$$
For $J_{2}$, the first four terms of (\ref{bot.6.f}) can be estimated by $(\ref{ml.1})_{2}$ and the last two terms of (\ref{bot.6.f}) can be estimated by $(\ref{ml.1})_{3}$, then
$$
\Big\|\int_{\R^{6\times 2}}J_{2}f_{0}d\Xi\Big\|_{L^{2}_{x}L^{2}_{\xi}}\leq e^{-\frac{2c_{0}t}{3}}\|f_{0}\|_{L^{2}_{x}H^{2}_{\xi}}\,.
$$
This means
\begin{align*}
\|\pa^{2}_{x}\mathbb{M}^{t}_{2}f_{0}\|_{L^{2}_{x}L^{2}_{\xi}}\leq e^{-\frac{2c_{0}t}{3}}\|f_{0}\|_{L^{2}_{x}H^{2}_{\xi}}
\Big(\int_{\mathbb{T}}\frac{1}{s_{1}s_{3}}dS\Big)\leq t^{2}e^{-\frac{2c_{0}t}{3}}\|f_{0}\|_{L^{2}_{x}H^{2}_{\xi}}\,,
\end{align*}
this completes the proof of the lemma.
\qed
\section{Fluid Part and Remainder Part}\label{se2}
\subsection{Green function}
Consider the linearized Landau equation
\begin{equation}\label{bgk.2.a}
\left\{\begin{array}{l}\displaystyle\pa_{t}f+\xi\cdot\nabla_{x}f=Lf\,,
\\ \\
\displaystyle f(x,0,\xi)=I(x,\xi)\,,
\\
\end{array}
\right.\end{equation}
where $I$ satisfies the zero mean conditions {\rm(\ref{in.1.d})}.
Hereafter, we will use just one index to denote the 3-dimensional sums with respect to the vector $k=(k_{1},k_{2},k_{3})\in \mathbb{Z}^{3}$, hence we set
$$
\sum_{k\in\mathbb{Z}}=\sum_{(k_{1},k_{2},k_{3})\in\mathbb{Z}^{3}}\,.
$$
Consider the Fourier series of initial condition $I$ in $x$
\begin{equation}\label{bgk.2.b}
\left\{\begin{array}{l}
\displaystyle
I(x,\xi)=\sum_{k\in\mathbb{Z}}(\hat{I})_{k}(\xi)e^{i\pi\eps k\cdot x}\,,
\\ \\
\displaystyle(\hat{I})_{k}(\xi)=\frac{1}{|\T^{3}_{1/\eps}|}\int_{\T^{3}_{1/\eps}}I(\cdot,\xi)e^{-i\pi\eps k \cdot x}dx\,,
\\
\end{array}
\right.\end{equation}
rewrite the solution $f(x,t,\xi)$ of (\ref{bgk.2.a}) as Fourier series
\begin{equation}\label{bgk.2.c}
\left\{\begin{array}{l}
\displaystyle
f(x,t,\xi)=\sum_{k\in\mathbb{Z}}(\hat{f})_{k}(t,\xi)e^{i\pi\eps k\cdot x}\,,
\\ \\
\displaystyle(\hat{f})_{k}(t,\xi)=\frac{1}{|\T^{3}_{1/\eps}|}\int_{\T^{3}_{1/\eps}}f(\cdot,t,\xi)e^{-i\pi\eps k\cdot x}dx\,,
\\
\end{array}
\right.\end{equation}
the Fourier modes of (\ref{bgk.2.b})--(\ref{bgk.2.c}) satisfy the following equations
\begin{equation}
\left\{\begin{array}{l}
\displaystyle
\pa_{t}\hat{f}_{k}+i\pi\eps\xi\cdot k \hat{f}_{k}-L\hat{f}_{k}=0\,,
\\ \\
\displaystyle \hat{f}_{k}(0,\xi)=(\hat{I})_{k}\,.
\\
\end{array}
\right.\end{equation}
Hence
\begin{equation}\label{bgk.2.d}
\hat{f}_{k}(t,\xi)=e^{(-i\pi\eps\xi\cdot k +L)t}(\hat{I})_{k}(\xi)\,,
\end{equation}
the solution of (\ref{bgk.2.a}) is given by
\begin{align*}
  f(x,t,\xi)&=\sum_{k\in\mathbb{Z}}e^{i\pi\eps k\cdot x+(-i\pi\eps\xi\cdot k +L)t}(\hat{I})_{k}(\xi)\,, \\
  &=\sum_{k\in\mathbb{Z}}\frac{1}{|\T^{3}_{1/\eps}|}\int_{\T^{3}_{1/\eps}}e^{i\pi\eps k \cdot(x-y)+(-i\pi\eps\xi\cdot k +L)t}
I(y,\xi)dy\,,
\end{align*}
the Green function $\mathbb{G}_{\eps}^{t}$ can be expressed as
\begin{equation}\label{bgk.2.f}
\displaystyle
\mathbb{G}_{\eps}^{t}=\sum_{k\in\mathbb{Z}}\frac{1}{|\T^{3}_{1/\eps}|}e^{i\pi\eps k\cdot x+(-i\pi\eps\xi\cdot k +L)t}\,.
\end{equation}
Note that if $\eps\to 0$, the Green function $\mathbb{G}^{t}_{0}(x,\xi,t)$ becomes an integral formula
\begin{equation}
\displaystyle
\mathbb{G}^{t}_{0}=\int_{\R^{3}}e^{i\eta\cdot x+(-i\pi\xi\cdot\eta+L)t}d\eta\,.
\end{equation}

\subsection{Long Wave Short Wave Decomposition}
We can decompose the Green function (\ref{bgk.2.f}) into the long wave part $\mathbb{G}^{t}_{\eps, L}$ and the short wave part $\mathbb{G}^{t}_{\eps, S}$ given respectively by
\begin{equation}\begin{array}{l}\label{bot.2.e}
\displaystyle
\mathbb{G}^{t}_{\eps, L}(x,\xi)=\sum_{|\eps k|<\de}\frac{1}{|\T^{3}_{1/\eps}|}e^{i\pi\eps k\cdot x+(-i\pi\eps\xi\cdot k +L)t}\,,
\\ \\
\displaystyle
\mathbb{G}^{t}_{\eps, S}(x,\xi)=\sum_{|\eps k|>\de}\frac{1}{|\T^{3}_{1/\eps}|}e^{i\pi\eps k\cdot x+(-i\pi\eps\xi\cdot k +L)t}\,.
\end{array}
\end{equation}
The following long wave short wave analysis relies on spectral analysis (Proposition \ref{pr12}).
\begin{lemma}{\rm(Short wave $\mathbb{G}^{t}_{\eps, S}$)} For any $s>0$, $I\in H^{s}_{x}L^{2}_{\xi}$, we have
\begin{equation}\begin{array}{l}\label{bot.2.f}
\|\mathbb{G}^{t}_{\eps, S}I\|_{L^{2}_{x}L^{2}_{\xi}}\leq e^{-O(1)t}\|I\|_{L^{2}_{x}L^{2}_{\xi}}\,,
\\ \\
\|\mathbb{G}^{t}_{\eps, S}I\|_{H^{s}_{x}L^{2}_{\xi}}\leq e^{-O(1)t}\|I\|_{H^{s}_{x}L^{2}_{\xi}}\,.
\end{array}
\end{equation}
\end{lemma}
\noindent{\it Proof.} Note that
\begin{equation*}
\mathbb{G}^{t}_{\eps, S}I=\sum_{|\eps k|>\de}e^{(-i\pi\eps\xi\cdot k +L)t}e^{i\pi\eps k\cdot x}\hat{I}_{k}(\xi)\,.
\end{equation*}
For $L^{2}_{x}L^{2}_{\xi}$ estimate, we have
\begin{equation*}
\|\mathbb{G}^{t}_{\eps, S}I\|^{2}_{L^{2}_{x}}=\sum_{|\eps k|>\de}
|e^{(-i\pi\eps\xi\cdot k +L)t}\hat{I}_{k}(\xi)|^{2}\,,
\end{equation*}
and hence by spectrum property (Proposition \ref{pr12}), we obtain
\begin{align*}
 \|\mathbb{G}^{t}_{\eps, S}I\|^{2}_{L^{2}_{x}L^{2}_{\xi}}&=\sum_{|\eps k|>\de}
\|e^{(-i\pi\eps\xi\cdot k +L)t}\hat{I}_{k}(\xi)\|_{L^{2}_{\xi}}^{2}\leq e^{-O(1)t}
\sum_{|\eps k|>\de}
\|\hat{I}_{k}(\xi)\|_{L^{2}_{\xi}}^{2} \\
  &=e^{-O(1)t}\int_{\R^{3}}\sum_{|\eps k|>\de}|\hat{I}_{k}(\xi)|^{2}d\xi
\leq  e^{-O(1)t}\|I\|_{L^{2}_{x}L^{2}_{\xi}}^{2}\,.
\end{align*}
For the high order estimate, we have
\begin{equation*}\begin{array}{l}
\displaystyle \|\mathbb{G}^{t}_{\eps, S}I\|^{2}_{H^{s}_{x}}\leq\sum_{|\eps k|>\de}(1+|\pi\eps k|^{2} )^{s}
|e^{(-i\pi\eps\xi\cdot k +L)t}\hat{I}_{k}(\xi)|^{2}\,,
\end{array}\end{equation*}
and hence
\begin{equation*}\begin{array}{l}
\displaystyle \|\mathbb{G}^{t}_{\eps, S}I\|^{2}_{H^{s}_{x}L^{2}_{\xi}}\leq e^{-O(1)t}\sum_{|\eps k|>\de}
(1+|\pi\eps k|^{2} )^{s}\|\hat{I}_{k}\|_{L^{2}_{\xi}}^{2}
\leq e^{-O(1)t}\|I\|_{H^{s}_{x}L^{2}_{\xi}}^{2}\,.
\end{array}\end{equation*}
\qed

In order to study the long wave part $\mathbb{G}_{\eps, L}^{t}$, we need to decompose the long wave part as the fluid part and non-fluid part, i.e. $\mathbb{G}_{\eps, L}^{t}=\mathbb{G}_{\eps, F}^{t}+\mathbb{G}_{\eps, L;\perp}^{t}$, where
\begin{equation}\begin{array}{l}\label{bot.2.g}
\displaystyle \mathbb{G}_{\eps, F}I=\sum_{|\eps k|<\de}\sum_{j=0}^{4}e^{\sigma_{j}(-\eps k)t}e^{i\pi\eps k\cdot x}\big<e_{j}(\eps k ), \hat{I}_{k}\big>_{\xi}e_{j}(\eps k )\,,
\\ \\
\displaystyle \mathbb{G}^{t}_{\eps, L;\perp}I=\sum_{|\eps k|<\de}e^{i\pi\eps k\cdot x}\Pi_{\de}\hat{I}_{k}\,.
\end{array}\end{equation}
\begin{lemma}{\rm(Long wave $\mathbb{G}^{t}_{\eps, L}$)} For any $s>0$, $I\in L^{2}_{x}L^{2}_{\xi}$ and satisfying the zero mean condition {\rm(\ref{in.1.d})}, we have
\begin{equation}\begin{array}{l}\label{bot.2.h}
\|\mathbb{G}^{t}_{\eps, L;\perp}I\|_{H^{s}_{x}L^{2}_{\xi}}\leq e^{-O(1)t}\|I\|_{L^{2}_{x}L^{2}_{\xi}}\,,
\\ \\
\|\mathbb{G}_{\eps, F}I\|_{H^{s}_{x}L^{2}_{\xi}}\leq e^{-O(1)\eps^{2} t}\|I\|_{L^{2}_{x}L^{2}_{\xi}}\,.
\end{array}\end{equation}
\end{lemma}
\noindent{\it Proof.} For the non-fluid part, using the spectrum property (Proposition \ref{pr12}), we have
\begin{equation*}\begin{array}{l}
\displaystyle\|\mathbb{G}^{t}_{\eps, L;\perp}I\|^{2}_{H^{s}_{x}}\leq\sum_{|\eps k|<\de}\big(1+|\pi\eps k |^{2}\big)^{s}
|\Pi_{\de}\hat{I}_{k}(\xi)|^{2}
\leq\big(1+|\pi\de |^{2}\big)^{s}\sum_{|\eps k|<\de}
|\Pi_{\de}\hat{I}_{k}(\xi)|^{2}\,,
\end{array}\end{equation*}
and hence
\begin{equation*}\begin{array}{l}
\|\mathbb{G}^{t}_{\eps, L;\perp}I\|^{2}_{H^{s}_{x}L^{2}_{\xi}}
\leq e^{-O(1)t}\|I\|_{L^{2}_{x}L^{2}_{\xi}}^{2}\,.
\end{array}\end{equation*}
For the fluid part, by spectrum property (Proposition \ref{pr12}) and the zero mean conditions {\rm(\ref{in.1.d})}, we have
\begin{align*}
 \|\mathbb{G}_{\eps, F}I\|^{2}_{H^{s}_{x}L^{2}_{\xi}}&\leq\big(1+|\pi\eps k |^{2}\big)^{s}
\sum_{|\eps k|<\de}\sum_{j=0}^{4}|e^{\sigma_{j}(-\eps k)t}||\big<e_{j}(\eps k ), \hat{I}_{k}\big>_{\xi}|^{2} \\
 &\leq C
\sum_{|\eps k|<\de}\sum_{j=0}^{4}|e^{\sigma_{j}(-\eps k)t}|\| \hat{I}_{k}\|_{L^{2}_{\xi}}^{2}\\
&\leq C
\sum_{|\eps k|<\de}\sum_{j=0}^{4}e^{-a_{j,2}|k\eps|^{2}[1+O(\de^{2})]t}\| \hat{I}_{k}\|_{L^{2}_{\xi}}^{2}\\
&\leq C
e^{-O(1)\eps^{2}t}\| I\|_{L^{2}_{x}L^{2}_{\xi}}^{2}\,.
\end{align*}
\qed

$\mathbf{Remark}$
(i) If $\eps>\de$, we do not have long wave part, i.e. $\mathbb{G}^{t}_{\eps, L}=0$.

\noindent(ii) The high order estimate of short wave part requires regularity in $x$. This is because $|\pi\eps k|$ may not be bounded. One needs the regularity of $I$ to ensure the decay of $\mathbb{G}^{t}_{\eps, S}$ in time.

\noindent(iii) In order to remove the regularity assumption in $x$, we need the Picard-type iteration for constructing the increasingly regular kinetic-like waves in section \ref{se4}.
\begin{theorem}
For any $I\in H_{x}^{2}L_{\xi}^{2}$ and satisfies the zero mean conditions {\rm(\ref{in.1.d})}, we have the following exponential time decay estimate about the linearized Landau equation {\rm(\ref{bgk.2.a})}
\begin{equation*}
\|\mathbb{G}_{\eps}^{t}I\|_{L^{\infty}_{x}L^{2}_{\xi}}\leq e^{-\la_{S}t}\|I\|_{H^{2}_{x}L^{2}_{\xi}}
+e^{-\la_{L}t}\|I\|_{L^{2}_{x}L^{2}_{\xi}}\,.
\end{equation*}
\noindent{\rm(i)} If $\eps>\de$, then $\la_{S}=O(1)$ and $\la_{L}=\infty$,
\\
\noindent{\rm(ii)} If $\eps\leq\de$, then $\la_{S}=O(1)$ and $\la_{L}=O(1)\eps^{2}$.
\end{theorem}
\subsection{Improvement of Fluid Part}\label{se3}
In this subsection, we improve the estimate of the fluid part. Recall the fluid part of the Landau equation (\ref{bot.2.g})
\begin{equation*}
\mathbb{G}^{t}_{\eps, F}I=\sum_{|\eps k|<\de}\sum_{j=0}^{4}e^{\sigma_{j}(-\eps k)t}e^{i\pi\eps k\cdot x}\big<e_{j}(\eps k ), \hat{I}_{k}\big>_{\xi}e_{j}(\eps k )\,.
\end{equation*}
We have
\begin{align*}
  \|\mathbb{G}_{\eps, F}^{t}I\|^{2}_{L^{2}_{\xi}}&=\sum_{j=0}^{4}\Big|\sum_{|\eps k|<\de}e^{\sigma_{j}(-\eps k)t}e^{i\pi\eps k\cdot x}\big<e_{j}(\eps k ), \hat{I}_{k}\big>_{\xi}\Big|^{2} \\
 &=\sum_{j=0}^{4}\Big|\frac{1}{|\T^{3}_{1/\eps}|}\sum_{|\eps k|<\de}\int_{\T^{3}_{1/\eps}}\int_{\R^{3}}e^{\sigma_{j}(-\eps k)t}e^{i\pi\eps k\cdot (x-y)}e_{j}(\eps k )I(y,\xi)d\xi dy\Big|^{2}\,.
\end{align*}
Note that
$$
\frac{1}{|\T^{3}_{1/\eps}|}\sum_{|\eps k|<\de}e^{-(\eps |k|)^{2}t}=\frac{1}{t^{3/2}}\frac{1}{|\T^{3}_{1/\eps}|}\sum_{|\eps k|<\de t^{1/2}}e^{-(\eps |k|)^{2}}
\to \frac{1}{t^{3/2}}\int_{B_{\de t^{1/2}}(0)}e^{-y^{2}}dy
$$
as $\eps\to 0$, this means for any $\alpha_{0}>0$ there exists $\de_{0}>0$ such that if $\eps<\de_{0}$, then
$$
\frac{1}{|\T^{3}_{1/\eps}|}\sum_{|\eps k|<\de t^{1/2}}e^{-(\eps |k|)^{2}}<\alpha_{0}+ \int_{\R^{3}}e^{-y^{2}}dy
\equiv C_{0}\,.
$$
Hence
\begin{align*}
  \|\mathbb{G}_{\eps, F}^{t}I\|_{L^{2}_{\xi}}&\leq
\Big(\frac{1}{|\T^{3}_{1/\eps}|}\sum_{|\eps k|<\de}e^{-(\eps |k|)^{2}t}\Big)\int_{\T^{3}_{1/\eps}}\|I\|_{L^{2}_{\xi}}dx \\
 &\leq O(1)
\Big(\frac{1}{|\T^{3}_{1/\eps}|}\sum_{|\eps k|<\de}e^{-(\eps |k|)^{2}t}\Big)e^{-\eps^{2}t}\\
&\leq O(1)
\frac{C_{0}}{(1+t)^{3/2}}e^{-\eps^{2}t}\,.
\end{align*}
\begin{theorem}
Assume that $\eps<\de$, $I\in L_{\xi}^{2}$ with compact support in $x$ and satisfies the zero mean conditions {\rm(\ref{in.1.d})}, then there exist $C_{0}, \de_{0}>0$ such that if $0<\eps<\de_{0}$, then
\begin{equation*}
\displaystyle\|\mathbb{G}_{\eps, F}^{t}I\|_{L^{2}_{\xi}}\leq  O(1)
\frac{C_{0}}{(1+t)^{3/2}}e^{-\eps^{2}t} \,.
\end{equation*}
\end{theorem}
If $\eps\to 0$, i.e. the whole space case, the pointwise estimate of the fluid part becomes
$$
\|\mathbb{G}_{0, F}^{t}I\|_{L^{2}_{\xi}}\leq O(1)\frac{1}{(1+t)^{3/2}}\,.
$$
This recover the whole space result in \cite{[LiuYu]}.
\subsection{Remainder Part}
The remainder part can be defined as follows:
\begin{equation}\label{bgk.4.n}
\mathbb{G}_{\eps, R}^{t}I=\mathcal{R}-\mathbb{G}_{\eps, F}^{t}I\,,
\end{equation}
combining long wave-short wave decomposition and kinetic decomposition, we have
\begin{equation}\label{bgk.4.o}
\mathbb{G}_{\eps, R}^{t}I=\mathbb{G}^{t}_{\eps, S}I+\mathbb{G}^{t}_{\eps, L;\perp}I-\sum_{j=-1}^{4}h^{(j)}=\mathcal{R}-\mathbb{G}_{\eps, F}^{t}I\,,
\end{equation}
hence by (\ref{bot.2.f}), (\ref{bot.2.h}) and (\ref{bgk.4.kq}),
\begin{equation}\label{bgk.4.p}
\big\|\mathbb{G}_{\eps, R}^{t}I\big\|_{L^{2}_{x}L^{2}_{\xi}}\leq t^{5}e^{-O(1)t}\|I\|_{L^{2}_{x}L^{2}_{\xi}}\,.
\end{equation}
This means $\mathbb{G}_{\eps, R}^{t}I$ is in non-fluid part, hence
\begin{equation}\label{bgk.4.o11}
\frac{d}{dt}\|\pa^{2}_{x}\mathbb{G}_{\eps, R}^{t}I\|_{L^{2}_{x}L^{2}_{\xi}}\leq-c\|\pa^{2}_{x}\mathbb{G}_{\eps, R}^{t}I\|_{L^{2}_{x}L^{2}_{\xi}}
+\|K(\pa^{2}_{x}h^{(4)})(\cdot,t)\|_{L^{2}_{x}L^{2}_{\xi}}\,.
\end{equation}
We only need to estimate
\begin{equation}\label{bgk.4.q}
\int_{0}^{t}\|K(\pa^{2}_{x}h^{(4)})(\cdot,t-s)\|_{L^{2}_{x}L^{2}_{\xi}} ds=O(1)\,.
\end{equation}
This immediately from (\ref{in.2.aa}) and (\ref{bgk.4.v}),
\begin{align*}
 \int_{0}^{t}\|K(\pa^{i}_{x}h^{(2i)})(\cdot,t-s)\|_{L^{2}_{x}L^{2}_{\xi}} ds
&\leq\int_{0}^{t}
\|K\|_{L^{2}_{x}L^{2}_{\xi}}\|\pa^{i}_{x}h^{(2i)}\|_{L^{2}_{x}L^{2}_{\xi}}ds\\
 &\leq\|I\|_{L^{2}_{x}L^{2}_{\xi}}\int_{0}^{t}(t-s)^{i+1}e^{-\frac{c_{0}(t-s)}{2}}ds
\\
&=O(1)\|I\|_{L^{2}_{x}L^{2}_{\xi}}\,.
\end{align*}
Hence we have
\begin{equation}\label{bgk.4.x}
\big\|\mathbb{G}_{\eps, R}^{t}I\big\|_{H^{2}_{x}L^{2}_{\xi}}\leq O(1)e^{-O(1)t}\,,
\end{equation}
moreover, by Sobolev inequality, we have
\begin{equation}
\big\|\mathbb{G}_{\eps, R}^{t}I\big\|_{L^{\infty}_{x}L^{2}_{\xi}}\leq O(1)e^{-O(1)t}\,.
\end{equation}

\end{document}